\documentclass{aastex}
\usepackage{emulateapj5}

\newcommand{\EGRET}{{\it EGRET}\ }

\newcommand{\et}{{\it et~al.}}

\newenvironment{inlinefigure}{
\smallskip
\def\@captype{figure}
\noindent\begin{minipage}{0.999\linewidth}\begin{center}}
{\end{center}\end{minipage}\smallskip}

\slugcomment{\apj \, submitted}
\shorttitle{Northern $\gamma$-ray Blazars}
\shortauthors{Sowards-Emmerd, Romani \& Michelson}

\begin{document}

\title{The Gamma-Ray Blazar Content of the Northern Sky}

\author{David Sowards-Emmerd\altaffilmark{1}, Roger W. Romani \& Peter F. Michelson\altaffilmark{1}}
\affil{Department of Physics, Stanford University, Stanford, CA 94305-4060}
\altaffiltext{1}{also, Stanford Linear Accelerator Center, Stanford, CA, 94039-4349}
\email{dse@darkmatter.stanford.edu, rwr@astro.stanford.edu, peterm@stanford.edu}

\begin{abstract}
	Using survey data, we have re-evaluated the correlation of 
flat spectrum radio sources with \EGRET sources in the Northern sky.
An analysis incorporating the radio and X-ray properties
and the $\gamma$-ray source localization is used to gauge the
reliability of associations and to search for counterparts of previously
unidentified \EGRET sources. Above $\;\vert b \vert$=10$^\circ$, where the 
classification is complete, 
we find that 70\% of the Northern \EGRET sources have counterparts similar 
to the bright \EGRET blazars. For several of these we identify known blazar
counterparts more likely than the earlier proposed 3EG association; for $\sim$20
we have new identifications.  Spectroscopic confirmation of 
these candidates is in progress and we have found flat spectrum radio 
quasars and BL Lac counterparts with redshifts as high as 4. We also 
find strong evidence for a set of 28 objects with no plausible counterpart 
like the known \EGRET blazars.  These thus represent either a new 
extragalactic population or a population of Galactic objects 
with a large scale height. The survey has been extended into the plane, where 
we find several new blazar candidates; the bulk of the sources are,
however, Galactic.
        Looking ahead to the GLAST era, we predict that several of 
the present 3EG sources are composite and that higher resolution data will 
break these into multiple blazar IDs.  
\end{abstract}


\section{Introduction}

	The \EGRET telescope on the {\it CGRO} satellite has detected
271 sources in a survey of the Gamma-ray ($\sim 100\,$MeV to 10\,GeV) sky.
Of these roughly a quarter have been identified as blazars and along the
Galactic plane there are a half dozen objects confirmed as young
pulsars through their pulsed $\gamma$-ray emission. Thus the bulk 
of the sources remain to be identified. There is a young Galactic population
along the plane clearly correlated with high mass stars \citep{kc96, yr97}.
An intermediate latitude
excess, especially in the direction of the Galactic bulge, suggests
an older Galactic population whose nature is uncertain. Finally there
remain a substantial number of high latitude sources with no AGN
identification.

	The selection of blazar candidates has largely proceeded by
correlation with existing radio surveys \citep{har99, mat01}.
In contrast, the Galactic plane sources and individual intermediate
latitude sources have been the subject of targeted multi-wavelength campaigns
\citep{rrk01, hal01, wal02}. In this project
we attempt to obtain a more complete census of plausible blazar counterparts,
sifting sources with extant radio survey data and then conducting
a multi-wavelength follow-up. We are obtaining spectroscopic confirmation
of the candidate AGN with Hobby$\bullet$Eberly
Telescope (HET) Marcario LRS spectroscopy. This survey has already discovered 
a number of new likely $\gamma$-ray blazars, including good candidates for 
the most distant persistent $\gamma$-ray sources known.

\subsection{Blazar Properties}

	The `blazar' label is somewhat heterogeneous, but in the context
of the unified AGN model, these sources are believed to be viewed close
to the axis of a powerful
relativistic jet. As such they are compact flat spectrum radio
sources, with apparent superluminal motion at VLBI scales. The
optical counterparts exhibit significant polarization and OVV (optically 
violently variable) behavior \citep{up95}. Optical spectroscopy often shows
large equivalent width emission lines (flat spectrum radio quasars) while
a significant fraction are continuum-dominated BL Lac-type objects. The 
broad-band spectral energy distribution (SED) is sometimes used to divide 
these into two classes with `red' blazars showing a synchrotron peak in 
the IR-optical with a Compton peak in the $\gamma$-ray while
`blue' blazars have a synchrotron component extending into the X-ray with
a Compton peak inferred to extend to the TeV range (Urry 1999, and references
therein).
There are surveys underway to substantially increase the number of
known blazars (e.g. DXRBS; Landt \et \,2001) which will eventually help 
in systematizing the broad-band properties of these sources.

	Of these properties bright, flat spectrum radio emission seems
best correlated with $\gamma$-ray activity, but this may be primarily
a selection effect of present counterpart lists, which focused on the
brightest radio sources as the principal candidates.  In an effort to make
a less biased census of the counterparts, we have re-examined the correlation
with X-ray and radio properties in selecting candidates.

\section{Counterpart Selection and `Figure of Merit' Ranking}

	The existing $\gamma$-ray blazar lists were largely selected from
the Green Bank 6cm and 21cm single dish surveys \citep{con91,wb92}.
With relatively poor resolution, confusion and extended jet/host galaxy emission
were a serious impediment to selecting flat spectrum cores as blazar
counterparts. This was especially true at lower radio fluxes and toward
the Galactic plane. Accordingly few sub-Jy counterparts have been identified.
Real progress can now be made since interferometric surveys covering much
of the Northern sky are available at 21cm (NVSS; Condon, \et\, 1988) and 
3.5cm (CLASS; Myers, \et\, 2002 --
the source list was kindly supplied in advance of publication by Ian Browne).
The CLASS survey, targeting compact gravitational
lens candidates, preselected flat spectrum sources by comparing NVSS 21cm
and Green Bank single dish 6cm fluxes, mapping sources with spectral index 
$\alpha \le 0.5$ ($S_\nu \propto \nu^{-\alpha}$) with the VLA A array, to
resolve structure at the 0.2$^{\prime\prime}$ scale. Coverage was complete
for Dec$\ge 0^\circ$ and $|b| \ge 10^\circ$; there was also partial coverage 
beyond these limits.  We are primarily interested here in the isolated
unresolved cores that are not lens candidates.

	By comparing the 3.5cm and 21cm interferometric fluxes we have 
a greatly improved
estimate of the spectral index of the compact cores, which are well identified 
by sub-arcsecond positional matches.  We find that a number
of sources previously selected as `flat' have extended high frequency emission.
Of course with such a variable population, non-simultaneous observations
can have erroneous spectral indices, but our estimates are certainly much
better than extrapolations from the confused GB6 fluxes.

	When CLASS coverage is not available, we do need to use GB6 fluxes.
Following the CLASS prescription, we subdivide the 6cm flux between the 
coincident 21cm NVSS sources to make spectral index estimates. We then
extrapolate with the 4.85GHz/1.4GHz spectral index to 8.4GHz to estimate a flux
in the CLASS range. In general, this extension was needed in the Galactic
plane where a plethora of 21cm sources make spectral indices unreliable.
However, in a number of cases we are able to show that no target meets
the CLASS survey criterion in the \EGRET error box. This will be used to rule
out typical blazar-type counterparts. All sources classified as blazars
ended up having a CLASS detection, except 3EG J2016+3657.

	We augment this radio selection by using the convenient, relatively
uniform RASS bright and faint source catalogs. Counterpart candidates are
evaluated for suitability of optical follow-up with USNO-B1.0/POSS photometry.
Our analysis will use the positions of these counterparts in the 
`Test Statistic' (TS) maps of the Third \EGRET (3EG) catalog, which plot the 
estimated likelihood of the observed $\gamma$-rays as a function 
of point source location.

\subsection{Radio/X-ray properties}

	High frequency radio emission is clearly a good discriminant. However
to make the least biased selection of counterparts, we compute the 
over-density of sources near high latitude ($\; \vert b \vert > 20^\circ$)
gamma-ray detections in bins of radio flux, 
spectral index and x-ray flux. To do this, we compare the number of sources
detected within a given TS probability contour with many random realizations
of the sky. The latter were obtained by shifting the \EGRET TS maps on the 
sky in 2 degree increments, computing the random source TS values and correcting
for variation in sky area from the survey cuts. The source counts within a given
confidence contour as a function
of, e.g., 8.4GHz flux, were then compared to the random realizations after 
normalization by the effective sky coverage. We define the excess fractional
source density associated with the 3EG source as
\begin{equation}
n = \frac{N_{3EG} - N_{Random}}{N_{3EG}}
\end{equation}
for each flux bin. The distribution for 8.4 GHz flux within the
95\% confidence region is shown
in figure 1. A simple least-squares power law fit to the binned data then
defines our excess source density function for evaluation of individual
candidates.
This exercise shows significant excess at fluxes well below the $\sim$Jy
limit considered in earlier analyses. Other authors have also
recently concluded that fainter radio sources should be considered as plausible
candidates \citep{wal02}.

\begin{inlinefigure}
\figurenum{1}
\scalebox{1.}{\rotatebox{0}{\plotone{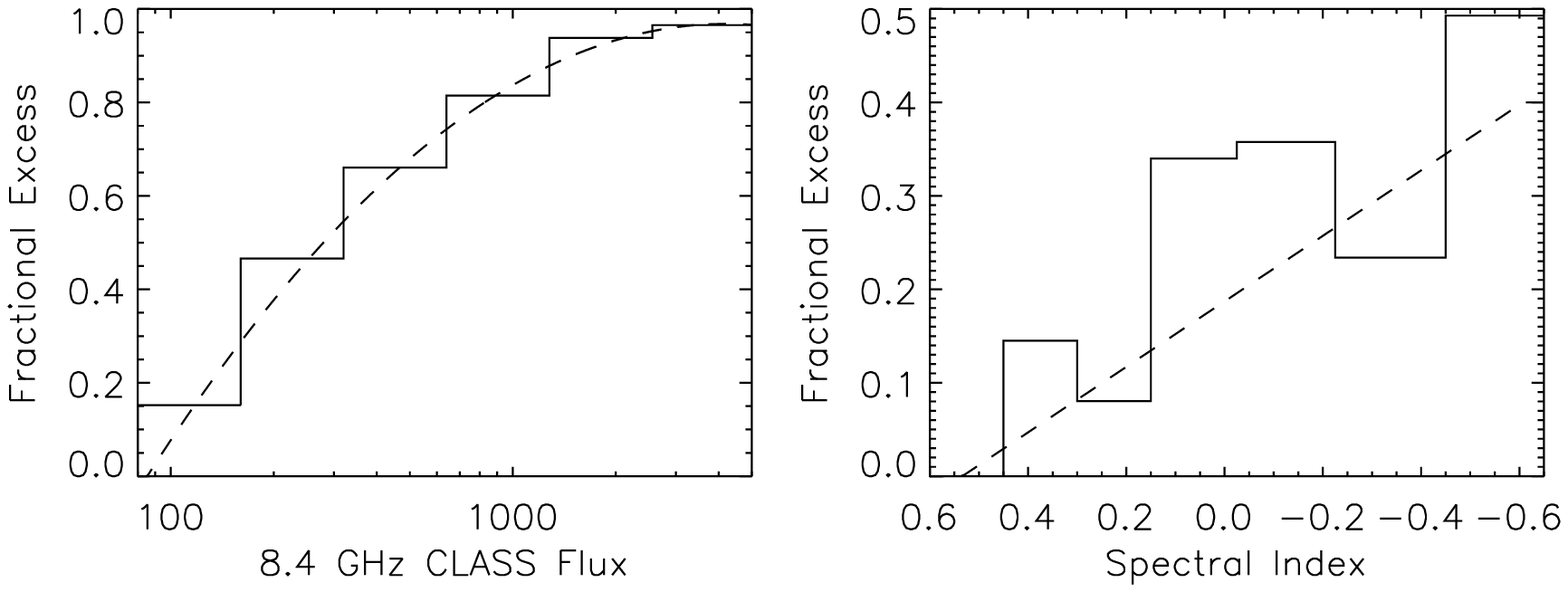}}}
\figcaption{The excess fractional source density within
\EGRET 95\% error contours as a function of: 8.4 GHz (CLASS)
flux (left), and 1.4GHz/8.4GHz spectral index (right).}
\label{radprob}
\end{inlinefigure}

We have similarly measured the excess of flat spectrum sources as a function of
spectral index. Here the correlation is strongest for $\alpha \la -0.6$,
becoming negligible for $\alpha \ge 0.53$. Correlation with the RASS
source catalog also shows an excess. However the maximum excess for the 
brightest sources is only 0.5 and about half of the strong
radio counterpart candidates were not detected in the relatively shallow
all-sky survey (some are present in deeper pointed observations). Clearly
we do not wish X-ray non-detection to exclude a candidate, although X-ray
bright sources are somewhat more likely to be correlated with the \EGRET
detections. We thus add 0.5 to the best fit X-ray $n$ so that it 
runs from 0.5 to 1.0.  In this way X-ray detection can increase the
source's selection by up to a factor of 2 (for $>$1PSPC cps), but X-ray 
non-detection is not seriously constraining. Thus the X-ray component of our
analysis is not crucial -- if it is excluded no sources are lost from our final
source list, although their detailed ranking does change somewhat.
The excess probability functions exhibited 
no strong dependence on the gamma-ray source localization.  Therefore, the 
area within the 95\% contours were taken to be representative of the 
gamma-ray sky for the derivation of the merit functions. 

\subsection{Source Position Weighting}

	The combination of 8.4GHz flux, spectral index and X-ray flux provides
a useful tool for identifying \EGRET-like blazars. Indeed, we are studying
a set of objects selected by these cuts, without any 3EG coincidence,
as possible GLAST-detectable blazars. Given the strong $\gamma$-ray
variability and the limited duty cycle in the flaring state, it would not 
be very surprising for relatively bright sources to have
eluded detection during the limited \EGRET exposure. However, to select
{\it individual} \EGRET counterparts, we will also use the radio source
position in the likelihood analysis. Previous blazar identification efforts
have used the circular or elliptical fits to the TS maps as an approximation
to the allowed sky region. However, many TS maps contain open or complex 
contours at the 95\% level, so we employ instead the precise TS at the
radio source position.

	The value at our counterpart candidate position $(\alpha, \delta)$
is compared to the map maximum, giving $\Delta TS 
= TS_{max} - TS(\alpha,\delta)\;$. $\Delta TS$ has been calibrated
by Monte Carlo simulations \citep{mat96} to give the statistical probability 
$L[\Delta TS(\alpha, \delta)]$ that the source lies within the contour 
of constant $\Delta TS$ for 50\%, 68\%
95\% and 99\% confidence levels. We linearly interpolate between these values
to obtain a smooth $L[\Delta TS(\alpha, \delta)]$; sources at 
large $\Delta TS$ have little probability of association with the 3EG
source, even if very blazar-like.
The statistical error estimate for $\Delta TS$ does not, however,
account for position uncertainties associated with nearby unmodeled 
$\gamma$-ray sources and with other systematic and instrumental biases.
This is confirmed
by plotting the distribution of bright 3EG blazars \citep{mat01} as a 
function of $\Delta TS$.  Considerably fewer than half of the sources lie 
within the 50\% contour, and several lie outside the 99\% contour.  
We find that the additional spread can be well modeled by dividing
the map value of $\Delta TS$ by 1.5 and then computing the nominal
statistical positional probability. The probabilities renormalized in
this way are a good match to the observed source distribution.
%
%
\subsection{Combined Figure of merit}

	We combine the fractional excesses for each of the counterpart
candidate's properties with the estimate of the positional probability
to compute a total `Figure of Merit' for the source
as a radio/X-ray counterpart:
\begin{equation}
FoM = n_{8.4GHz} \times n_{\alpha} \times n_{x-ray} \times L(\alpha,\delta)
\end{equation}
Notice that this product of excess source fraction and positional uncertainty
is {\it not} a normalized probability of source identification. It is
however an unbiased ranking of the counterpart likelihood that can be
compared across the entire (high latitude) 3EG population. To avoid confusion, 
we multiply by $100$ in the quoted FoM values.

	To evaluate the significance of this FoM statistic, we produced
random realizations of the radio/X-ray sky by random, independent RA
and DEC draws from the true CLASS position list, after excluding the 
actual positions of the $\sim70$ highest-FoM coincident sources
[which would otherwise imprint the observed correlation on the 
random realizations]. We compute the average distribution of FoM
for $10^3$ realizations of such random skies, and compare with the true FoM
distribution in Figure 2. The hashed region shows the $\pm 1\sigma$ range 
for our estimate of the fraction of sources in a given FoM bin that are 
`real', i.e. in excess of random coincidence. For FoM$< 0.1$ there 
is little excess correlation. There is a puzzling deficit of true sources 
with FoM$\sim 2$, but we have not been
able to trace this to any one of the FoM source properties and so conclude
that this is a statistical fluctuation. We select sources with FoM$\ge 0.25$
(i.e. $< 20\%$ false positive, even for the lowest FoM) as good counterparts; 
this is about a factor
of two above the lowest bins showing significant correlation.  To facilitate 
comparison with earlier ID lists, we divide these sources in half with 
FoM$>1$ designated as `likely' counterparts and $0.25 < {\rm FoM} < 1.0$ 
as `plausible' counterparts (note however that our `plausible' sources 
have a relatively high fraction of true associations).
The line in figure 2 shows a simple linear fit to the source
probability with a strong decrease below FoM=0.1. Integrating through
the distribution of measured FoM, we find that of the 35 `likely'
sources we expect less than 3 false positives ($\ge 92\%$ good IDs)
and of the 32 `plausible' sources $\le 6$ may be false positives
($\ge 82\%$ good IDs). We can compare the FoM distributions of our IDs
with those designated in the 3EG catalog and in \citet{mat01}. Many of the 
highest FoM sources are common to all lists. However, both the 3EG and Mattox
lists claim high confidence for several sources that we only assign
low probability. In addition each list has a number of sources that are
not flat spectrum or compact and do not meet our interferometric selection
criteria. These sources, 11 in the 3EG and 7 in \citet{mat01} are mostly
lower confidence (`a' and `plausible' designations in these catalogs). 
However, in addition to selecting previous IDs, we also find $\sim 50\%$
more high confidence sources than either, and virtually all of our
`plausible' sources are new. Noting that \citet{mat01} retained
`plausible' sources with estimated likelihoods as low as a few percent,
while our lowest FoM sources have a likelihood $> 80\%$, we believe that
our list is less biased, more complete and more reliable than
earlier efforts. In particular, we can more easily identify
multiple associations with a given \EGRET source, rather than taking only the
`best' blazar ID.

\begin{inlinefigure}
\figurenum{2}
\scalebox{1.}{\rotatebox{0}{\plotone{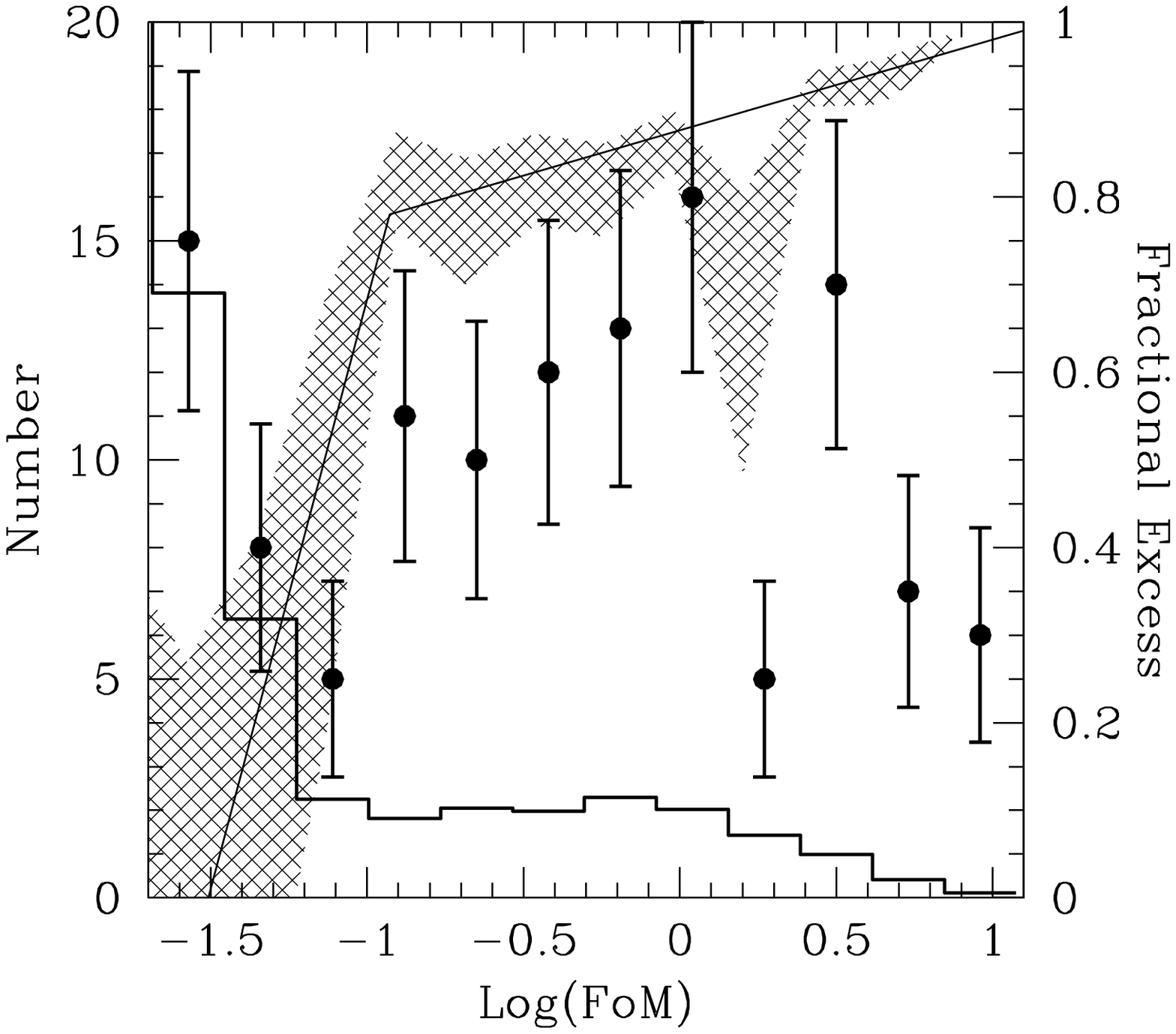}}}
\figcaption{Random (histogram) and true (points, with Poisson error bars)
distributions of our FoM. The fractional excess (true ID fraction) for each bin
is shown by the (Poisson) error range of the shaded region (right scale).
Note the rapid fall-off of reliable identification below Log(FoM)$\approx -0.9$.
}
\label{FoMfig}
\end{inlinefigure}

\subsection{Radio Counterpart IDs}

	We find at least one blazar candidate counterpart
for 66 of the Northern 3EG sources. In many cases, the previously
claimed 3EG `high confidence' and `plausible' blazars are recovered
with the largest FoM value for a given error ellipse. 
However, in a number of cases, our criteria select a different blazar
as much more likely than the claimed catalog association. In addition,
we find some 2 dozen new radio associations in the Northern sky, whose
FoM confidence are at least as good as that of the lowest
of the previously identified `high confidence' sources. In several cases
there are multiple radio blazar IDs in a single \EGRET error ellipse.
One rather complex example is 3EG J0118+0248. In the 3EG catalog this was 
possibly identified with 0119+041; our analysis does select it as a (low FoM)
possible association, but it is well outside the 99\% error contour.
In the \citet{mat01} analysis, the 3EG source was attributed to 3C37 ($z=0.672$); 
this AGN however has a quite steep spectrum for the compact component 
and does not meet our blazar identification criteria.  Instead, our
analysis selects two compact flat spectrum sources as the most likely
counterparts.  The first is a relatively low luminosity radio galaxy core
at z=0.047, the second is a newly discovered flat spectrum radio QSO at 
z=4.0. The 3EG error contours are quite elongated with the major axis spanning
these two sources (Figure 3), although 0119+041 is also a plausible source
of this extension.
A second example is 3EG J0808+5114, which
hosts a flat spectrum radio quasar at z=1.14 and a BL Lac at z=0.13,
again separated along the major axis of an extended $\gamma$-ray uncertainty
region.  Given that the 3EG survey is strongly flux
limited, that the $\gamma$-ray blazar luminosity function is quite steep,
and that the poor $\gamma$-ray resolution causes substantial source overlap
and confusion, it is not surprising that in some cases a combination
of fainter sources can push a location above the detection threshold.
In several other cases support for multiple IDs comes from complex or
elongated likelihood (`TS') maps.

\begin{inlinefigure}
\figurenum{3}
\scalebox{1.}{\rotatebox{0}{\plotone{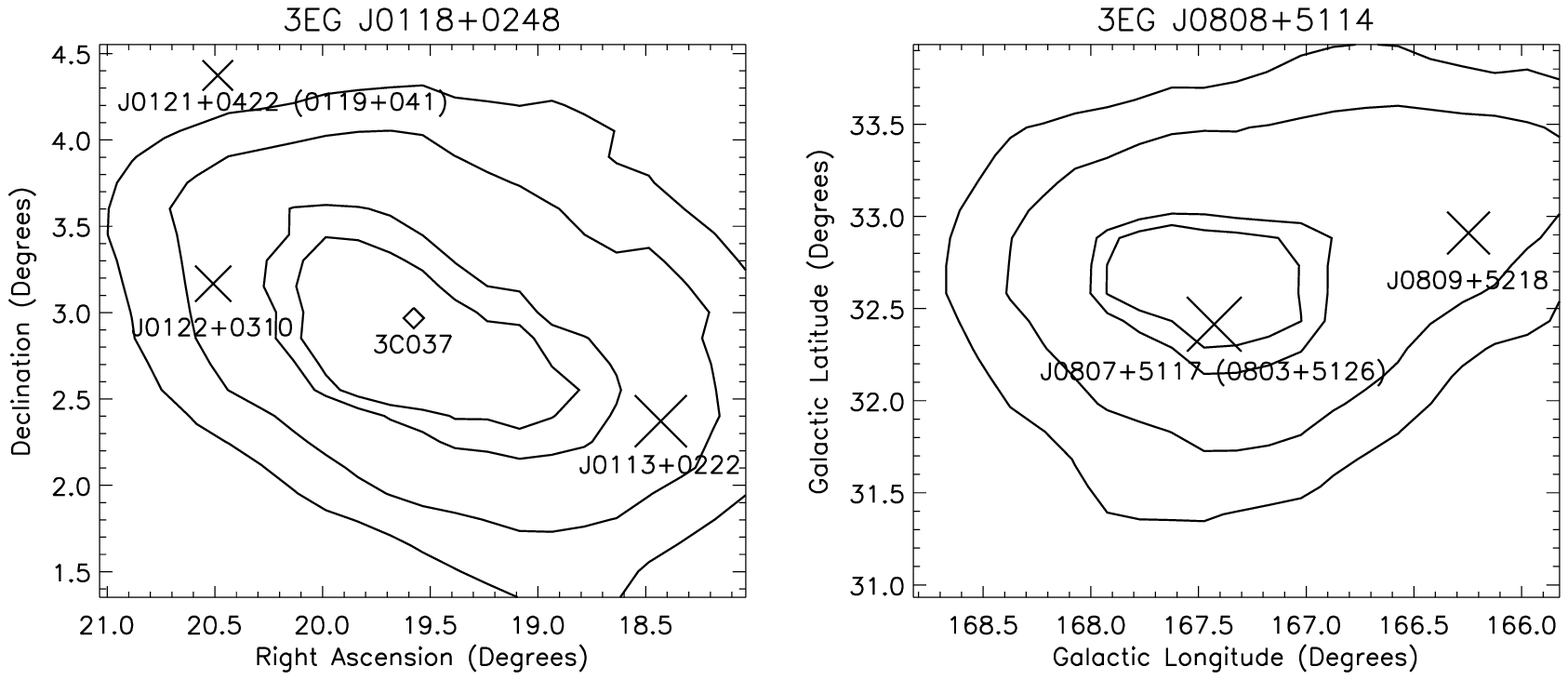}}}
\figcaption{TS maps of possible composite 3EG sources. Left 3EG J0118+0248. 
The 3EG identification 0119+041, the steep spectrum \citet{mat01} counterpart 
3C037 (diamond) and our two new blazar counterparts (along the uncertainty
region major axis) are shown.  Right: 3EG J0808+5114.
Again two high confidence identifications lie along the major axis.
}
\label{radprob}
\end{inlinefigure}

\subsection{Radio non-IDs}

	Since the GB6/NVSS/RASS provide a uniform survey (abetted by CLASS
follow-up), the absence of plausible IDs for some \EGRET sources is
significant. For example, we find no IDs for two high latitude
sources believed to be likely pulsars (3EG J0010+7300= CTA 1 and
3EG J1835+5918, see Reimer \et\, 2001). Our technique allows
us to select {\it individual} 3EG sources as non-blazar IDs. This is
particularly useful at intermediate latitude, as it lets us
separate the blazar fraction of the population, leaving a cleaner
non-blazar sample. North of Dec=0$^\circ$, 28 previously unidentified sources
are selected as {\it not} like \EGRET blazars. There is some concentration
to the Galactic plane, but much of this sample is at high latitude, 
suggesting either a new class of extragalactic $\gamma$-ray sources
or an old high scale height Galactic populations of pulsars or similar
objects. 

\subsection{Summary Results}

	We summarize our source classification in Table 1 and Figure 4. 
In summary, of the 116 Northern 3EG sources (excluding the Solar flare), 66 
have at least one plausible blazar-like radio counterpart. In the Table
our high confidence associations are listed in bold face and our lower
confidence associations in plain text. Previously claimed AGN associations
not supported by our analysis are given in italics in parentheses, with
FoM values computed when possible. The Table also notes 10 sources 
associated with pulsars or known plerion/young SNR (indented). These are
neutron star/pulsar/SNR associations from the 3EG catalog, with a few
additions from new discoveries which are noted in the individual source
comments. Of course many additional associations have been claimed for
Galactic objects; we do not evaluate or exhaustively review these here,
but we believe that beyond the few pulsed detections, the identifications
rapidly become quite speculative.

	An additional 28 sources are unidentified but according to our analysis
are definitively unlike the known 3EG blazars. The remaining 
sources are at low $|b| \le 5^\circ$ latitude. Only 
faint flat spectrum sources are allowed in the error boxes of 7 of these
and, given the high density of Galactic sources, these are almost certainly
chance coincidence. The final 5 show bright extended HII regions which
could, in principal, hide a bright flat spectrum compact blazar, but again
in practice we expect the sources to be Galactic. Further radio imaging
study could better rule out AGN counterparts in these cases.

	Figure 4 summarizes the present 3EG source
classifications in Galactic coordinates. Above the Dec=0 line, we use our new 
classifications from Table 1, with blazars as filled circles.
In the South, the same symbols are used for the 3EG
`A' and `a' blazar designations. Pulsar IDs and strong
pulsar/plerion candidates are shown by filled and open stars. 
The definitive non-blazars are shown by open circles.
The improved completeness of the Northern classification is evident.
Clearly pushing south into the bulge population will be very interesting.

	Since AGN are expected to be isotropic, it is of interest to
examine the distribution of the northern blazar IDs. Detailed
comparison requires folding through the \EGRET exposure and sensitivity maps
for a model blazar luminosity function; we defer this to a future communication.
However since exposure is relatively uniform for $|b|>20^\circ$, we report
simple number counts. From $30^\circ<|b|< 90^\circ$ we find 36 blazar 
identifications, or 11.6/sr within our survey boundary. Extrapolating toward 
the plane, we find for $20^\circ<|b|< 30^\circ$ 12 sources against 11 expected.
At $10^\circ<|b|< 20^\circ$ we find 8 sources vs. 12 expected, and a similar
fraction within $10^\circ$ of the plane. Given the higher
background and detection threshold near the plane, this does not however
mean that our identification of blazars at low $|b|$ is incomplete.
\vfill
\goodbreak

\section{Optical Follow-up}

Of our radio-selected counterparts, 50 are new sources, not selected in
previous \EGRET counterpart lists. Further, several additional selected 
radio sources, while flagged as possible counterparts in earlier studies, have
not had spectroscopic confirmation of their AGN nature. We first checked for
published spectroscopy of these radio sources, comparing with the 
Tenth Edition of the Quasar 
Catalog \citep{ver01} and cross checking by querying NED. We find that 
16 (5 high probability IDs) of our newly selected counterparts
have known redshifts. However over thirty sources in the Northern
sky were found to have previously unstudied blazar-like radio/X-ray
candidate counterparts, many with quite high FoM. These are the target of 
our HET spectroscopy, along with a number of similar sources having no
\EGRET association.

	Just before submission, we learned that \citet{hem03} have recently
completed follow-up spectroscopy of a set of previously claimed radio
counterparts of 3EG (and 2EG, GEV) sources with lower 4.8GHz fluxes.  Their 
list includes 8 sources also identified in our exercises. For three sources
we have new spectroscopy in common; for two their new redshift agrees with
that which we obtain (below). We were also able to obtain a redshift
for the third source (J1826+0149). For three other sources \citet{hem03}
obtain no $z$, for two the $z$ they measure was already found in the QSO
catalog, and for the last (J1239+0443) they obtain a new $z$. This value
is included in our table as the only entry not from cataloged data or our new
spectroscopy.

\subsection{HET LRS Spectroscopy}

Spectroscopy was obtained using the Marcario Low Resolution Spectrograph 
(LRS)\citep{hil98} on the 9.2m HET \citep{ram98}. This novel telescope, 
of tilted optical Arecibo design is still
under development, but with a range of magnitudes the blazar
candidates were suitable targets for the early operations phase. These targets
were observed in regular queue observations from 3/02 -2/03.
We obtained 2$\times$450s exposures for most targets; a few of the
most interesting fainter targets were observed with $2 \times 900$s
exposure. Observations were made employing a 300 l/mm grating and a
2$^{\prime\prime}$ slit for a dispersion of 4$\AA$ per (binned) pixel
and an effective resolution of 16$\AA$ covering $\lambda\lambda 4200-10500$.
The new spectra are displayed in Figure 5.

	Standard IRAF CCD reductions, optimal extraction and calibration
were performed.  Unfortunately, the varying pupil during the HET 
tracks is not yet fully monitored and so small (few \%) flat-field 
variations were not corrected, as with untracked flat field frames
the varying illumination worsened the
(already substantial) fringe features and sky subtraction errors in the near
IR. Redshifts were measured by cross correlation analysis with AGN and
galactic templates, using the IRAF RVSAO package. A few identifications
are based only on the MgII $\lambda$2798 line, but in each case the proposed $z$
is the only plausible value, given the absence of other expected strong lines.
Several of our objects 
are continuum dominated with very faint line features, for these the 
substantial light grasp of the HET was important. All of the observed targets
were found to be AGN, although some spectra were taken under rather 
poor conditions. Some of these have relatively uncertain $z$ estimates,
these are noted by (:) in Table 2.

\begin{inlinefigure}
\figurenum{6}
\scalebox{1.}{\rotatebox{0}{\plotone{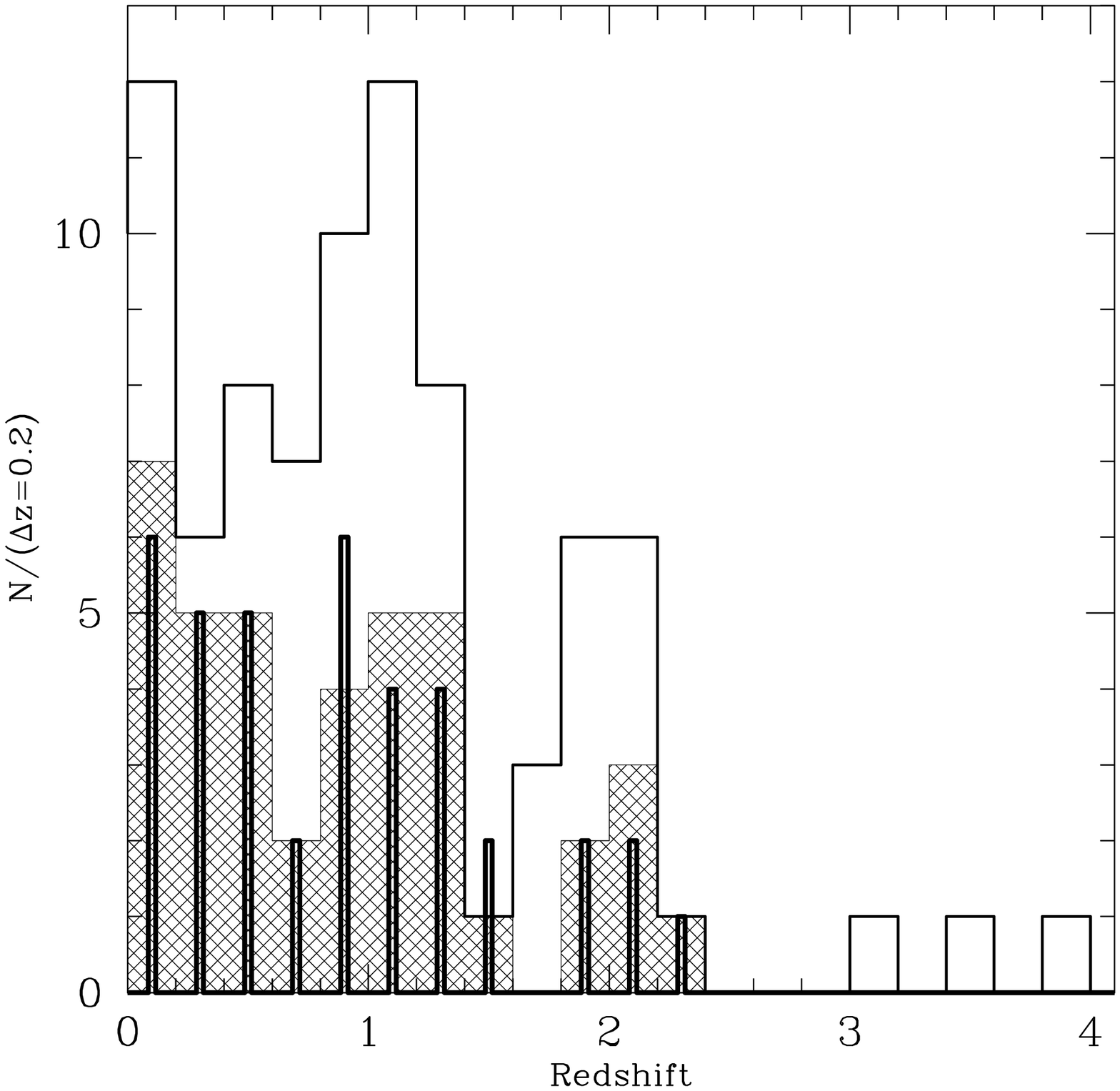}}}
\figcaption{The redshift distributions of our Northern blazar sample
(solid line histogram), the Northern identifications from the 3EG catalog 
(shaded histogram) and the \citet{mat01} sample (bar histogram). 
}
\label{FoMfig}
\end{inlinefigure}

	As of this publication we have obtained 29 new spectroscopic 
identifications and an additional 16 new associations with previous
spectroscopy.  The new and archival redshifts for
our new blazar IDs are given in Tables 1 and 2.  The later table also
contains precise positions and sky survey flux estimates; as expected for 
blazars, the flux at our observation epoch was often substantially different.
Most of the AGN were found to be flat spectrum radio quasars. We denote
sources with emission line equivalent width $\la 5\AA$ as BL Lacs.
Several sources have the narrow lines (FWHM$<1000$km/s) and lower 
excitation states characteristic
of radio galaxies, we denote these as `R' in Table 1. It should be noted
that within the classical AGN paradigm, narrow-line radio galaxies would 
not be expected to show blazar activity.  The redshift distribution
of our identified sources is shown in Figure 6. For comparison, we also
plot the redshift distributions of the sources selected by the 3EG and 
\citet{mat01} criteria.  The $z$ distributions are rather
similar, largely displaying spectroscopic ID selection effects, but it
is clear that we have substantially higher sensitivity to high redshift and
have nearly doubled the maximum $z$ value.

\section{Notes on Individual Sources}

\textbf{3EG J0118+0248} As noted above, the \citet{mat01} identification with 
the bright steep spectrum source
3C37 is not supported by our analysis and the 3EG association with 0119+041, is 
apparently superseded (or at least augmented) by our discovery of two
flat spectrum candidates along the error ellipse major axis. If these contribute
similarly to the $\gamma$-ray flux, they would represent extrema of the
$\gamma$-ray blazar population, with a factor of $\sim 10^4$ difference in 
luminosity.  If the association is confirmed, J0122+0310 at z=4.0 would be 
the highest redshift non-GRB $\gamma$-ray source known. This 3EG
source will be particularly interesting for study in the GLAST era, since 
with higher angular resolution the contributions of the several posited
associations can be disentangled and GLAST's good sensitivity above 10GeV 
should allow intergalactic absorption of J0122+0310 against a cosmic
background of starlight to be detected at $\ga 20$GeV.

\textbf{3EG J0204+1458}
In addition to the likely association J0204+1514 this source may also 
incorporate flux from the flatter and more centrally placed radio
source J0205+1444.

\textbf{3EG J0215+1123} with a reasonably large FoM J0213+1213 is an
interesting low $z$ BL Lac candidate for TeV studies.

\textbf{3EG J0222+4253}
This object has been argued to be a composite by \citet{kui00} with a possible
pulsed detection of the millisecond pulsar PSR J0218+42 accounting
for part of the sub-GeV flux. The remainder was attributed (also in the 3EG 
catalog) to the BL Lac 3C 66A. This source, with $\alpha_{1.4/8.4} = 0.62$, 
lies just below the spectral cut-off for inclusion in our FoM. However,
2.3\,GHz VLBA data are available \citep{fey00} which give a much flatter index
for the compact component, which we adopt in Table 1. This does make 
3C 66A highly likely.  CLASS analysis also selects the 215mJy source J0223+4259
(3C 66B, a radio galaxy with a core+jet at $z=0.02$) as a plausible counterpart. 
Confusion in the NVSS image prevents detection of the core, so using the 15GHz 
core flux measured by \citet{jac93} we obtain $\alpha_{8.4/15}=0.38$ and a strong
plausible association. The GeV localization of \citet{kui00} supports
either 3C object as the source of the higher energy $\gamma$-rays.
Certainly GLAST observations will be helpful sorting out this complex
region. Interestingly, a TeV detection is claimed in this region \citet{nes98}
which would be somewhat surprising from $z$ as large as 0.44. The apparent
position is consistent with either 3C 66A or 3C 66B and future air Cerenkov 
observations may be able to select the preferred source, given their 
$\sim 0.1^\circ$ separation.

\textbf{3EG J0245+1758}
The source closest to the TS maximum is a strong-lined FSRQ at z=3.59. This
source shows a likely damped Ly$\alpha$ system at $z\approx 3.15$, raising
the possibility that the source is gravitationally magnified by an
intervening Galaxy. 

\textbf{3EG J0329+2149}
\citet{mat01} noted the radio source as a possible association. We confirm 
it as a likely blazar counterpart and have measured z=2.07, in agreement
with \citet{hem03}.

\textbf{3EG J0404+0700}
The better of the two likely associations is our most puzzling case,
spectroscopically. When first observed, the spectrum was highly
continuum dominated.  The two strongest lines were picked up as Ly$\alpha$ 
and CIV $\lambda 1449$ at $z=3.13$ in
cross-correlation analysis. Both are (local) $\ga 5\sigma$ detections; they have
rest equivalent widths of $0.44$\AA\, and $0.39$\AA, respectively. 
CIII $\lambda 1909$
is obscured due to the fringing and no other strong lines are expected. This
redshift ID is supported by an apparent strong MgII absorption doublet 
at $z=1.578$ (REW 0.19/0.22$\AA$). With such low EW for the emission lines
this source is (in this state) clearly a BL Lac \citep{mar96}. However a second 
observation several months later caught the source in a low state for which 
broad MgII line and several Ne and O lines give a highly significant 
measurement of $z=1.133$. The MgII line is in fact also present in the high
state at identical flux. The difference spectrum shows no MgII line,
but does have the higher $z$ features noted above. We conservatively
adopt the lower redshift, as $z=3.13$ more than doubles that of the
highest $z$ BL Lac cataloged to date (PKS 1309-216 $z=1.49$ Blades, Murdoch 
\& Hunstead 1980). Further spectroscopy in a variety of flux states is
certainly warranted.

\textbf{3EG J0450+1105}
\citet{hem03} have observed J0449+1121, identified in the 3EG catalog as
the likely counterpart, and have not confirmed the catalog $z$. Their nearly
featureless spectrum suggests a BL Lac.

\textbf{3EG J0459+3352}
This Galactic error region has at least two maxima. The blazar
candidate J0503+3404 is well associated with one peak, but the second
peak, nearly 2 degrees away is more likely Galactic.

\textbf{3EG J0634+0521}
This source has been plausibly associated with the Be X-ray binary 
SAX J0635+0533
\citep{kar01}, so it might be classified as `p'.

\textbf{3EG J0808+5114}
This source is probably composite with two likely counterparts (Figure 3).


\textbf{3EG J0917+4427} The bright, 1.3Jy, flat spectrum association is only
plausible, being just outside the 99\% contour. However, this error region is
highly elongated and source is offset in the short direction. If an unmodeled
systematic of $\sim 0.3^\circ$ induces this offset, this would be a likely 
association.

\textbf{3EG J1133+0033} and \textbf{3EG J1329+1708} we have observed a plausible
radio source in each, identifying them as featureless BL Lacs, but have not 
obtained redshifts.

\textbf{3EG J1605+1553}
Our new identification at z=0.11 is brighter at 8.4GHz and has
a smaller $\Delta TS$ than the association proposed by \citet{mat01}.

\textbf{3EG J1621+8203}
\citet{muk02} have previously argued for identification with this Seyfert.



\textbf{3EG J1835+5918}
This source has been the subject of intense study \citep{rei01,hal02}, who have
both argued that it is an isolated Geminga-like pulsar. 

\textbf{3EG J2016+3657}
This crowded Galactic error region contains the SNR CTB 87, several bright HII
regions in addition to the radio source B2013+370, claimed as a possible
BL Lac \citep{hal01}. Our spectrum of this source, taken under very poor 
conditions, is sufficient to support the continuum-dominated nature
of the heavily absorbed counterpart, but is insufficient to determine
a redshift. We tentatively adopt the blazar designation of these authors,
but confirmation requires improved angular resolution and possibly variability
correlation with GLAST.

\textbf{3EG J2021+3651} Is likely associated with the newly discovered 
pulsar and plerion discussed in \citet{rob02,rrk01}.

\textbf{3EG J2035+4441} May be associated with an X-ray/radio plerion
\citep{rrk01}.

\textbf{3EG J2036+1132} Our analysis produces FoM=0.22 for B2032+107
(a $z=0.6$ BL Lac), the 3EG A designated counterpart of this source. This 
is just below our cutoff, but as it is significantly less than the FoM of
J2034+1154, we propose re-assignment to this source. Our second source
J2031+1219 is bright and flat, but lies significantly outside the 99\%
confidence contour. A re-evaluation of the likelihood contours in terms 
of multiple sources could affect the rankings of these plausible counterparts.

\textbf{3EG J2206+6602} The 3EG catalog also selected this source with low
probability; we confirm it as a $z=1.12$ FSRQ.

\textbf{3EG J2209+2401}
This source was claimed as an identification in \citep{mat01}, but the
listed redshift was evidently a typographical error, repeating that
of another source. We have measured a single line $z$ that excludes the 
published result.

\textbf{3EG J2227+6122}
This source is likely identified with the recently discovered
X-ray/radio pulsar PSR J2229+6114
and its associated wind nebula \citep{hal01b}.

\section{Conclusions and Implications}

	We believe that we have made a relatively unbiased assessment of
the association of flat spectrum radio sources with \EGRET AGN, identifying
plausible counterparts down to 8.4GHz fluxes of $\sim 100$mJy.  Assuming,
as we have argued, that the bulk of our likely and plausible IDs are correct,
we have substantially increased the completeness of identification of the 
Northern $|b|\ge 10^\circ$ \EGRET sources from $\sim$40\% 
to $\sim 70$\%. We have also argued that in a number of instances the 3EG 
sources are composites; including this, the increase in the number of proposed
$\gamma$-ray blazars is even larger. 

	Our identification allows selection of sources with substantially
smaller 8.4GHz radio flux. It should not be too surprising that the number of 
candidates tapers off smoothly below the previous typical 1Jy value. We
suspect that with smaller GLAST error ellipses or improved multi-wavelength
constraints, identification could continue well below the $\sim 0.1$Jy
limit of our survey. Our identification of $\sim 30$ sources that have
flat spectrum counterparts (if any) well below 0.1Jy does not, of course, 
preclude that some of these may be radio-faint or steep spectrum AGN. Indeed, 
excess coincidences do continue slightly below our FoM identification
limit (Figure 2). Also a number of the 3EG and \citet{mat01} identifications
not duplicated here are steep spectrum, but very bright low $z$ AGN. In these
cases (with the exception of 3EG J0118+0248/3C037, discussed above) the steep 
spectrum proposed counterpart is located at large $\Delta TS$. Nevertheless,
if such sources are truly associated, it is reasonable that these form a 
distinct subset of the 3EG population, not identified in this analysis.
However, the fact that an appreciable fraction of the 
`non-blazar' 3EG sources correlate with the Galactic disk
suggests that many represent a new (old) Galactic population. Detailed
assessment of the completeness of blazar IDs through the plane requires
careful treatment of the 3EG exposure maps; we defer this and other population
analysis to a later paper.

	Along with fainter radio associations, we have pushed
back the horizon of plausible blazar identifications (Figure 4).
It will be interesting to compare the SED and VLBI $\beta_\perp$
properties of these fainter blazar counterpart candidates.
This survey has found several individual targets of particular interest, 
including good $\gamma$-ray source associations at $z=3-4$, several distant
BL Lacs, and several low
$z$ AGN that may be suitable targets for ground-based TeV observations.
These targets should be the blazars brightest to GLAST and they
will provide the most detailed spectra and light curves; as such they
merit further study in preparation for the GLAST era.
Since spectroscopic identifications are continuing, we defer detailed
discussion of the red shift/ luminosity function distributions to a future 
paper.

The `non-blazars' are also excellent targets for further study in preparation
for GLAST, since this sample will most likely produce new classes of
high energy emitters.  While we do not discuss here the correlation of 
$\gamma$-ray spectra and variability with the lower energy SEDs, we do
note that our `non-blazars' correlate fairly well with the `steady'
sources of \citet{get00}  and the `persistent' sources of
Grenier (2000), although exceptions occur. Clearly, pushing the improved
identifications south to cover the Galactic bulge and beyond will 
be very important for characterization of these populations.

\acknowledgments

	We thank Roger Blandford for helpful discussions on the nature of 
$\gamma$-ray blazar sources and for assistance with access to the CLASS
source list. Ian Browne and Thomas York are also thanked for passing along
an early copy of these important data. The referee, R.C. Hartman, and S. Digel
provided careful readings and comments that substantially improved the
manuscript.

We are particularly grateful to the HET team, especially
Gary Hill for bringing the Marcario LRS into operation and
Matt Shetrone, Jeff Mader, Brian Roman and the rest of the 
telescope team for getting these early observations done. We 
dedicate these HET results to
the late Stanford provost Gerald J. Lieberman, whose early support of
Stanford's participation in the HET made this project possible.

  The Hobby-Eberly Telescope is operated by McDonald Observatory on behalf 
of The University of Texas at Austin, the Pennsylvania State University, 
Stanford University, Ludwig-Maximilians-Universit\"at M\"unchen, and 
Georg-August-Universit\"at G\"ottingen.
The Marcario Low Resolution Spectrograph is a joint project of the 
Hobby-Eberly Telescope partnership and the Instituto de Astronomia de la 
Universidad Nacional Autonoma de Mexico.

This research has made use of the NASA/IPAC Extragalactic Database (NED) 
which is operated by the Jet Propulsion Laboratory, California Institute 
of Technology, under contract with the National Aeronautics and Space 
Administration. DSE was supported by SLAC under DOE contract
DE-AC03-76SF00515 and PFM acknowledges support from NASA contract
NAS5-00147.


\begin{inlinefigure}
\figurenum{4}
\scalebox{1.22}{\rotatebox{90}{\plotone{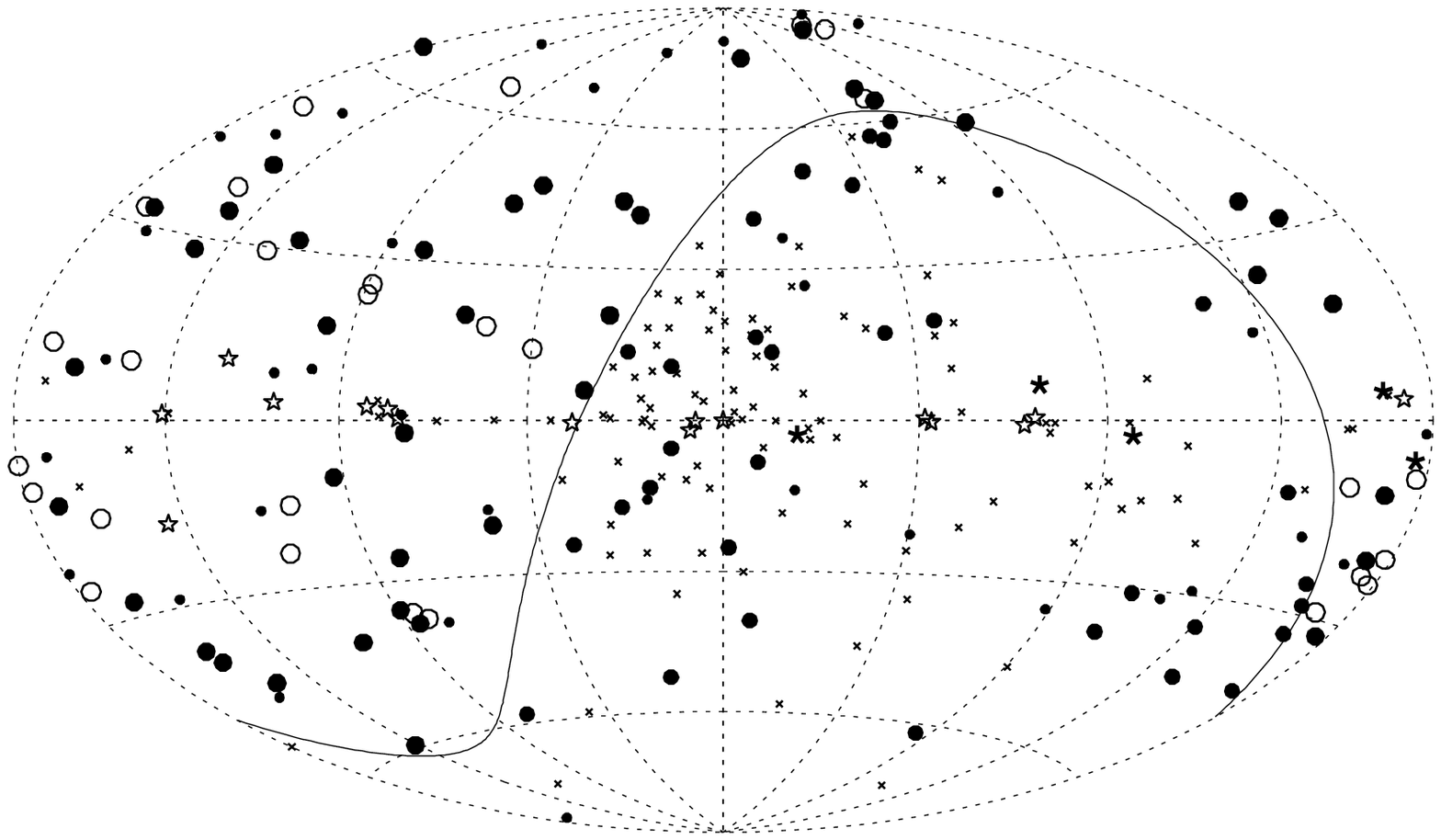}}}
\figcaption{Aitoff equal area projection of 3EG
sources in Galactic coordinates, showing our new
classifications in the Northern hemisphere. 
Large filled circle=high confidence blazar,
Smaller filled circle=plausible blazar, 
Filled star=pulsar, 
Open star=pulsar/plerion candidate,
Open circle= Non-Blazar,
cross=presently unclassified.
Symbols south of DEC=0$^\circ$ are similar, with AGN drawn from the 3EG A/a
classifications.
\label{Ait}}
\end{inlinefigure}

\begin{figure}
\figurenum{5}
\plotone{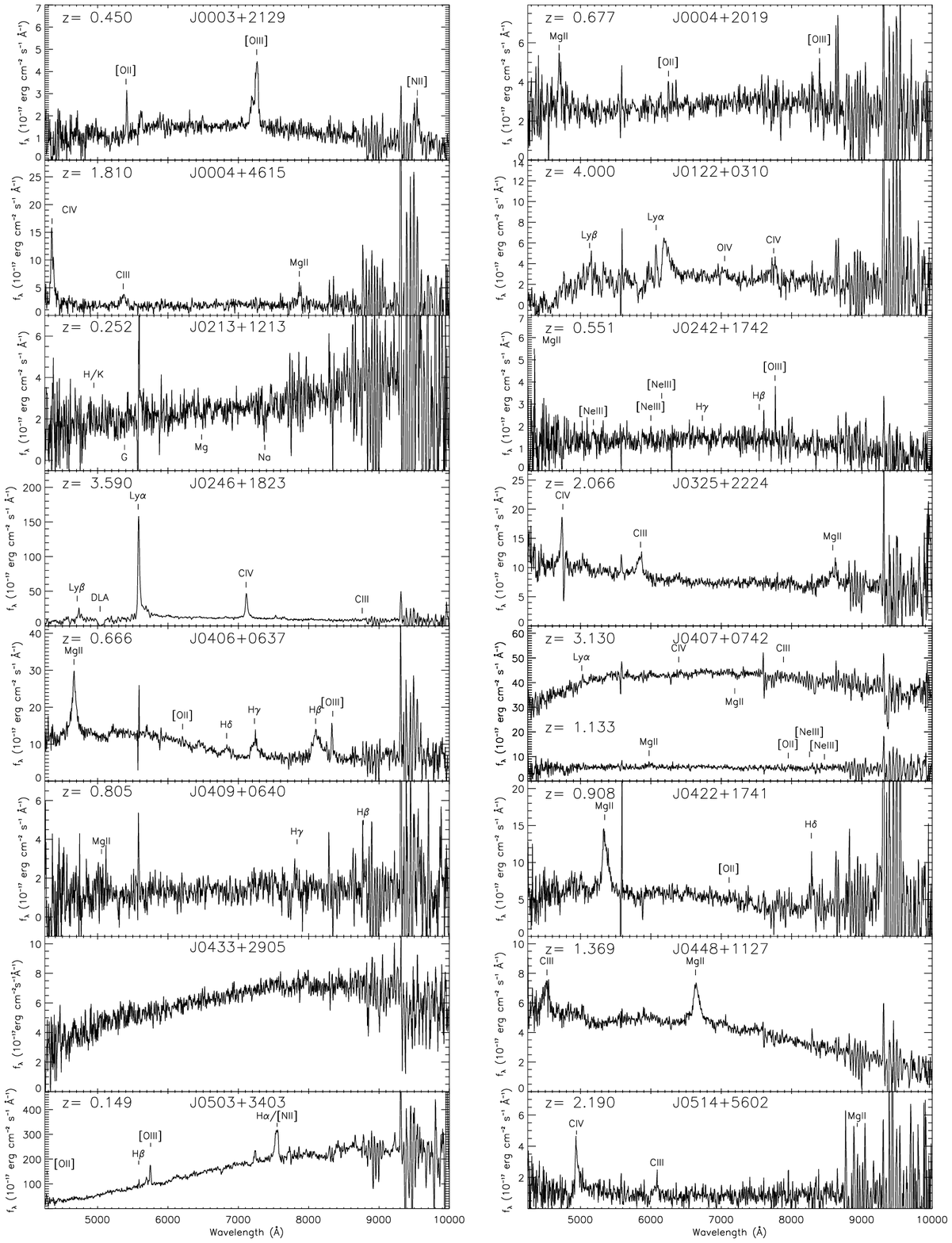}
\figcaption{HET/Marcario LRS  spectra.}
\label{spec}
\end{figure}
\begin{figure}
\figurenum{5}
\plotone{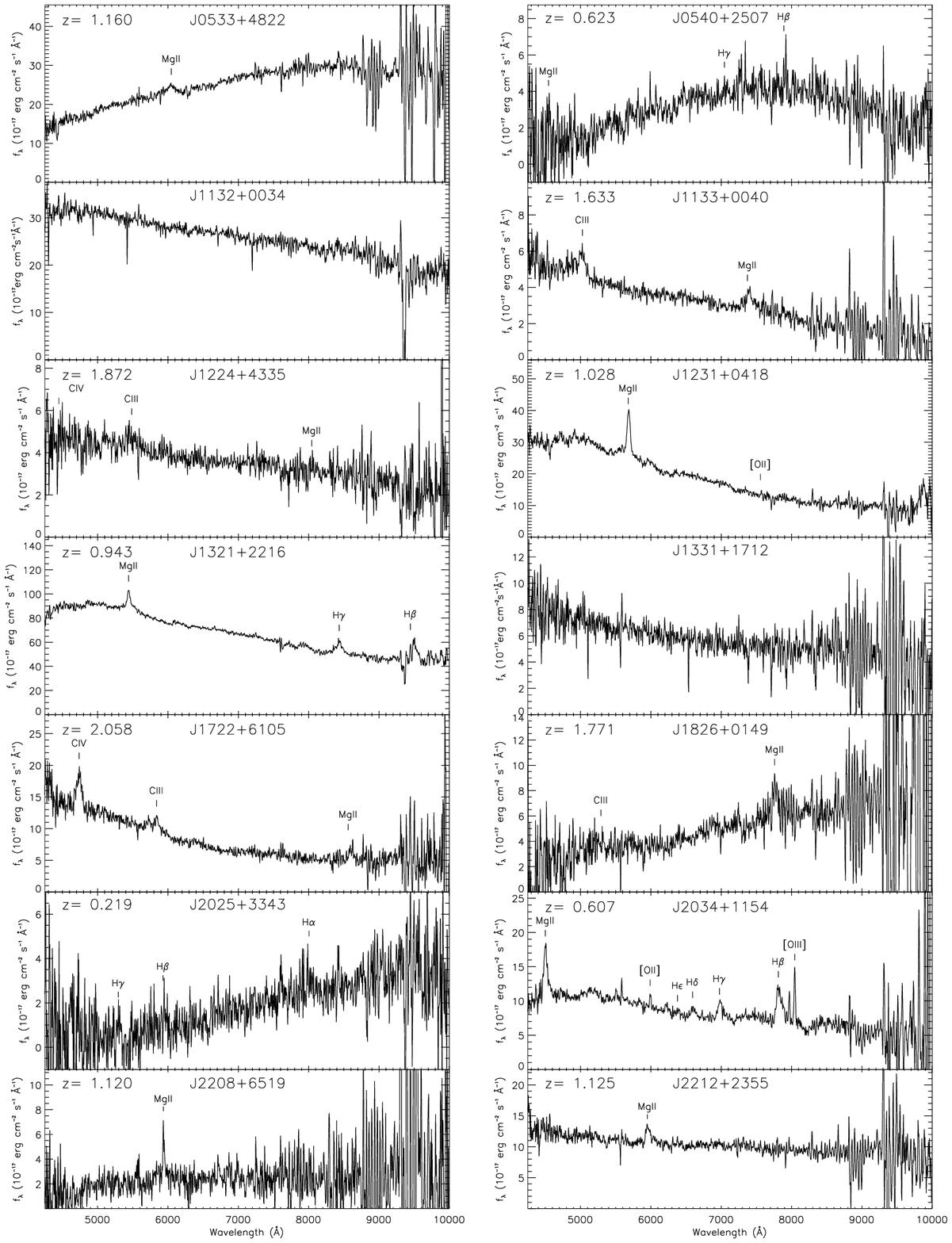}
\figcaption{(continued) HET/Marcario LRS  spectra.}
\label{spec}
\end{figure}


\clearpage
\begin{deluxetable}{|l|l|r|l|l|c|l|l|l|l|}
\tabletypesize{\scriptsize}
\tablecaption{Northern 3EG Objects}

\tablecomments{
Column 7:  Our new associations and/or redshifts * = new spectral ID. 
$\dagger$= redshift from NED, 10th QSO Catalog, etc.\\ 
Column 8: \citet{mat01} selected blazars +=`High probability', $-$=`Plausible'\\ 
Column 9: 3rd \EGRET Catalog blazars, A=`High confidence', a=` lower confidence'\\
Column 10: classification f=FSRQ, b=BL Lac, R=narrow-line radio galaxy,
G=Likely galactic, N='Non-blazar', p=Pulsar candidate/plerion, P=confirmed pulsar.}

\tablewidth{0pt}
\tablehead{
\colhead{3EG} & \colhead{ID} & \colhead{$S_{8.4}$} & \colhead{$\alpha$} & \colhead{FoM}& \colhead{z} & \multicolumn{4}{c}{Notes} }
\startdata
\tableline
J0010+7309
&\quad \quad CTA1&&&&& &&& pN\\
\tableline
J0118+0248
&\textbf{J0113+0222}&644&-0.1&3.55&0.05& $\dagger$&&&R\\
& J0122+0310&121&-0.1&0.59&4.00& *&&&f\\
& J0121+0422,0119+041&1351&-0.1&0.29&0.63&&&a&f\\ 
&\textit{(0115+027,3C037)}&--&&&0.67&&--&&\\
\tableline
J0204+1458
&\textbf{J0204+1514,0202+149}&3325&0.09&5.77&0.41& &+&A&f\\
&\textbf{J0205+1444}&185&0.01&2.58&---& *&&&\\
\tableline
J0215+1123
& J0213+1213&164&-0.1&0.53&0.25&  *&&&b\\
\tableline
J0222+4253
&\quad \quad PSR J0218+4232&&&&& &&& pb?\\
&\textbf{J0222+4302,0219+428,3C66A}&728&0.17&3.16&0.44& &+&A&b\\ 
& J0223+4259,3C 66B&215&0.38&0.90&0.02&$\dagger$&&& R\\ 
\tableline
J0229+6151
&&&&&&&&&G\\
\tableline
J0237+1635
&\textbf{J0238+1636,0235+164}&5453&-0.5&4.33&0.94&&+ &A&f\\
\tableline
J0239+2815
& J0237+2848,0234+285&3123&-0.1&0.58&1.21& &+&A&f\\
\tableline
J0241+6103
&\quad \quad LSI +61 303&&&&&&&&pN\\
\tableline
J0245+1758
&\textbf{J0242+1742}&236&-0.1&1.64&0.55& *&&&f\\
& J0246+1823&125&0.29&0.42&3.59& *&&&f\\
\tableline
J0323+5122
&&&&&&&&&G\\
\tableline
J0329+2149
&\textbf{J0325+2224,0322+222}&528&-0.0&1.40&2.07&*&--&&f \\
\tableline
J0348+3510
&&&&&&&&&N \\
\tableline
J0404+0700
&\textbf{J0406+0637}&227&-0.0&1.11&0.67&*&&&f \\  
&\textbf{J0407+0742}&524&-0.3&3.01&1.13&*&&&b \\
& J0409+0640&181&-0.1&0.93&0.81&* &&&f\\
\tableline
J0407+1710
&&&&&&&&&N\\
\tableline
J0416+3650
&&&&&&&&&G \\
&\textit{(0415+379,3C111)}&--&&&0.05&&&a&\\
\tableline
J0423+1707
& J0422+1741&131&-0.5&0.78&0.91& *&&&f\\
\tableline
J0426+1333
&&&&&&&&&N \\
\tableline
J0429+0337
&&&&&&&&&N\\
\tableline
J0433+2908
&\textbf{J0433+2905,0430+2859}&432&-0.0&6.15&---&&--&A&b\\
\tableline
J0435+6137
&&&&&&&&&N \\
\tableline
J0439+1555
&&&&&&&&&N\\
\tableline
J0439+1105
&&&&&&&&&N \\
\tableline
J0450+1105
& J0448+1127&206&0.17&0.54&1.37&*&&&f \\
&\textbf{J0449+1121,0446+112}&1226&-0.2&3.58&1.2?&&--&A&?\\
\tableline
J0459+0544
& J0457+0645&434&0.07&0.30&0.41&$\dagger$&&& f\\
& J0502+0609,0459+060&543&0.28&0.85&1.11&&--&A&f\\
& J0505+0459&808&0.10&0.81&0.95&$\dagger$ &&&f\\
\tableline
J0459+3352
& J0503+3403&448&0.38&0.59&0.15& *&&&R?\\
\tableline
J0500+2529
&&&&&&&&&N\\
&\textit{(0459+252)}&--&&&0.28&&--&&R\\
\tableline
J0510+5545
& J0514+5602&229&0.18&0.41&2.19& *&&&f\\
\tableline
J0516+2320
&\quad \quad SolarFlare&&&&&&&&S\\
\tableline
J0520+2556
&&&&&&&&&N\\
\tableline
J0521+2147
&&&&&&&&&N \\
\tableline
J0530+1323
&\textbf{J0530+1331,0528+134}&3074&-0.3&6.36&2.07&&+&A&f\\
\tableline
J0533+4751
&\textbf{J0533+4822,0529+4820}&556&-0.1&4.23&1.16&*&--&&f\\
\tableline
J0534+2200
&\quad \quad Crab&&&&&&&&P\\
\tableline
J0542+2610
& J0540+2507&207&0.12&0.63&0.62&*&&&f\\
\tableline
J0546+3948
&&&&&&&&& \\
\tableline
J0556+0409
&&&&&&&&&N \\
\tableline
J0613+4201
&&&&&&&&&N \\
\tableline
J0617+2238
&\quad \quad IC443&&&&&&&&p\\
\tableline
J0628+1847
&&&&&&&&&G\\
\tableline
J0631+0642
&&&&&&&&&G\\
\tableline
J0633+1751
&\quad \quad Geminga&&&&&&&&P\\
\tableline
J0634+0521
&&&&&&&&&G\\
\tableline
J0721+7120
&\textbf{J0721+7120,0716+714}&594&0.09&7.67&---&&+&A&b\\
\tableline
J0737+1721
&\textbf{J0738+1742,0735+178}&2942&-0.1&12.58&0.42&&+&A&b\\
& J0739+1739&114&-0.2&0.63&---&* &&&\\
\tableline
J0743+5447
& J0742+5444,0738+5451&142&0.36&0.56&0.72&&--&A&f\\
\tableline
J0808+4844
&&&&&&&&&N\\
&\textit{(0804+499)}&880&0.12&0.10&1.43&&&a&f\\
&\textit{(0809+483,3C196)}&--&&&0.87&&--&a&\\
\tableline
J0808+5114
&\textbf{J0807+5117,0803+5126}&358&-0.4&8.32&1.14&&&a&f\\
&\textbf{J0809+5218}&154&0.08&1.24&0.14&$\dagger$&&& b\\
\tableline
J0828+0508
&\textbf{J0831+0429,0829+046}&1225&-0.0&3.39&0.18&&+&A&b\\
\tableline
J0829+2413
&\textbf{J0830+2410,0827+243}&713&0.00&6.41&0.94&&+&A&f\\
\tableline
J0845+7049
&\textbf{J0841+7053,0836+710}&1757&0.42&1.14&2.22&&+&A&f\\
\tableline
J0853+1941
&\textbf{J0854+2006,0851+202}&2997&-0.3&7.48&0.31&&+&A&b\\
\tableline
J0910+6556
&&&&&&&&&N \\
\tableline
J0917+4427
& J0920+4441,0917+449&1368&-0.1&0.49&2.18&&&a&f\\
\tableline
J0952+5501
& J0957+5522,0954+556&1498&0.39&0.49&0.90&&+&A&f\\
\tableline
J0958+6533
&\textbf{J0958+6533,0954+658}&1269&-0.3&13.14&0.37&&+&A&b\\
\tableline
J1009+4855
&&&&&&&&&N\\
&\textit{(1011+496)}&252&0.21&0.24&0.20&&&a&b\\
\tableline
J1052+5718
& J1058+5628,1055+567&189&0.09&0.42&0.14&&&a& b\\
\tableline
J1104+3809
&\textbf{J1104+3812,Mrk 421,1101+384}&631&-0.0&8.52&0.03&&+&A&b\\
\tableline
J1133+0033
& J1132+0034&213&0.42&0.71&---& *&&&b\\
& J1133+0015&119&0.35&0.42&1.17&$\dagger$&&& f\\
&\textbf{J1133+0040}&320&0.05&4.07&1.63& *&&&f\\
\tableline
J1200+2847
& J1159+2914,1156+295&1233&0.24&0.98&0.73&&+&A&f\\
\tableline
J1212+2304
&&&&&&&&&N\\
\tableline
J1222+2315
&&&&&&&&&N \\
\tableline
J1222+2841
& J1221+2813,1219+285&1217&-0.2&0.34&0.10&&--&A&b\\
\tableline
J1224+2118
&\textbf{J1224+2122,1222+216}&1073&0.34&1.97&0.44&&+&A&f\\
\tableline
J1227+4302
& J1221+4411&435&0.12&0.49&1.35&$\dagger$&&& f\\
& J1224+4335&220&0.26&0.91&1.87& *&&&f\\
& J1226+4340&145&0.10&0.95&---& *&&&\\
\tableline
J1229+0210
&\textbf{J1229+0203,3C273,1226+023}&41725&0.04&8.77&0.16&&+&A&f\\
\tableline
J1235+0233
&&&&&&&&&N\\
\tableline
J1236+0457
&\textbf{J1231+0418}&302&0.05&1.34&1.03& *&&&f\\
& J1239+0443,1237+0459&290&0.09&0.96&1.75&&&a&f\\
\tableline
J1308+8744
&&&&&&&&&N\\
\tableline
J1323+2200
& J1321+2216&323&-0.0&0.54&0.94&*&&&f \\
& J1322+2148&147&0.26&0.29&---&* &&&\\
& J1327+2210,1324+224&2107&-0.5&0.75&1.40&&&a&f\\
\tableline
J1329+1708
& J1331+1712&120&0.21&0.45&---&*&&&b \\ 
&\textbf{J1333+1649,1331+170}&483&-0.1&2.08&2.09&&--&A&f\\
\tableline
J1337+5029
&&&&&&&&&N\\
\tableline
J1347+2932
& J1343+2844&192&0.13&0.34&0.91&$\dagger$&&& f\\
\tableline
J1424+3734
& J1419+3821&775&-0.1&0.90&1.83&$\dagger$ &&&f\\
& J1420+3721&158&0.06&0.25&0.97&$\dagger$ &&&f\\
& J1421+3855&132&-0.2&0.46&0.49&$\dagger$ &&&f\\
& J1426+3625&613&-0.2&0.44&1.09&$\dagger$ &&&f\\
\tableline
J1605+1553
&\textbf{J1603+1554}&256&-0.5&4.46&0.11&$\dagger$ &&&f\\
&\textit{(1604+159)}&223.5&0.56&0.00&0.36&&--&a&b\\
\tableline
J1608+1055
&\textbf{J1608+1029,1606+106}&1805&-0.1&3.37&1.23&&+&A&f\\
\tableline
J1614+3424
&\textbf{J1613+3412,1611+343}&3042&0.13&2.14&1.40&&+&A&f\\
\tableline
J1621+8203
&\textbf{J1632+8232,NGC6251}&738&0.05&1.63&0.02&&&&R\\
\tableline
J1635+3813
&\textbf{J1635+3808,1633+382}&2448&0.04&3.94&1.81&&+&A&f\\
\tableline
J1727+0429
&\textbf{J1728+0427,1725+044}&622&0.04&6.30&0.29&&--&A&f\\  
\tableline
J1733+6017
& J1722+6105&203&-0.1&0.95&2.06&*&&&f \\
& J1724+6055&166&0.17&0.61&---&* &&&\\
\tableline
J1738+5203
&\textbf{J1740+5211,1739+522}&1318&-0.2&11.25&1.38&&+&A&f\\
\tableline
J1822+1641
&&&&&&&&&N \\
\tableline
J1824+3440
&\textbf{J1826+3431}&289&0.25&1.11&1.81&$\dagger$&&& f\\
\tableline
J1825+2854
&&&&&&&&&N \\
&\textit{(1829+2905)}&722&0.78&0.00&0.84&&--&&\\ 
\tableline
J1828+0142
&\textbf{J1826+0149}&725&-0.2&1.39&1.77& *&&&f\\
\tableline
J1835+5918
&&&&&&&&&N\\
\tableline
J1850+5903
&&&&&&&&&N\\
\tableline
J1856+0114
&\quad \quad W44/PSR 1853+01&&&&&&&&pN\\
\tableline
J1903+0550
&&&&&&&&&G\\
\tableline
J1928+1733
&&&&&&&&& \\
\tableline
J1958+2909
&&&&&&&&&G\\
\tableline
J1959+6342
&\textbf{J2006+6424,2005+6416}&958&-0.3&2.99&1.57&&--&&f\\
\tableline
J2016+3657
&&&&&&&&&b\\
\tableline
J2020+4017
&\quad \quad $\gamma\;$ Cygni&&&&&&&&p\\
\tableline
J2021+3716
&\quad \quad PSR2021+3651&&&&&&&&p\\
\tableline
J2022+4317
&&&&&&&&& \\
\tableline
J2027+3429
&\textbf{J2025+3343}&2728&-0.4&2.85&0.22& *&&&f\\ 
\tableline
J2033+4118
&&&&&&&&& \\
\tableline
J2035+4441
&\quad \quad RRK&&&&&&&&p\\
\tableline
J2036+1132
& J2031+1219&1178&-0.1&0.26&1.22&$\dagger$ &&&f\\
& J2034+1154&216&0.24&0.61&0.61& *&&&f\\
&\textit{(2032+107)}&463&0.46&0.22&0.60&&--&A&b\\
\tableline
J2046+0933
&\textbf{J2049+1003}&888&-0.6&3.39&---& *&&&\\
\tableline
J2100+6012
& J2102+6015&164&0.35&0.41&---& *&&&\\
&\textit{(2105+598)}&179&0.57&0.00&&&&a&\\
\tableline
J2202+4217
&\textbf{J2202+4216,BL Lac,2200+420}&3321&0.31&3.44&0.07&&+&A&b\\
\tableline
J2206+6602
& J2208+6519,2206+650&249&0.33&0.41&1.12& *&&a&f\\
\tableline
J2209+2401
&\textbf{J2212+2355,2209+236}&719&-0.1&4.11&1.13& *&+&A&f\\
\tableline
J2227+6122
&\quad \quad PSR J2229+6114&&&&&&&&p\\
\tableline
J2232+1147
& J2232+1143,CTA 102,2230+114&2923&0.48&0.92&1.04&&+&A&f\\
\tableline
J2243+1509
&&&&&&&&&N\\
\tableline
J2248+1745
&&&&&&&&&N\\
\tableline
J2254+1601
&\textbf{J2253+1608,3C 454.3,2251+158}&10380&0.10&8.57&0.86&&+&A&f\\
\tableline
J2255+1943
&\textbf{J2253+1942,2250+1926}&362&-0.1&4.46&0.28&&&a& f\\
\tableline
J2314+4426
&&&&&&&&&N\\
\tableline
J2352+3752
&&&&&&&&&N\\
&\textit{(2346+385)}&243&0.22&0.24&1.03&&&a&\\ 
\tableline
J2358+4604
& J0004+4615&214&-0.3&0.28&1.81&*&&&f \\
& J2354+4553,2351+456&990&0.34&0.67&1.99&&--&A&f\\
\tableline
J2359+2041
& J0001+1914&504&-0.3&0.29&3.10&$\dagger$&&&f\\
&\textbf{J0003+2129}&269&-0.6&1.69&0.45& *&&&f\\
&\textbf{J0004+2019}&162&-0.6&1.17&0.68& *&&&f\\
&\textbf{J2358+1955,2356+196}&558&0.10&1.44&1.07&&--&A&f\\
\tableline
\enddata
\end{deluxetable}


\begin{deluxetable}{lllrllllc}
\tabletypesize{\scriptsize}
\tablecaption{New 3EG Counterpart Candidates with Spectroscopic Identification}
\tablecomments{Some sources have been previously flagged (see Table 1), but 
are spectroscopically confirmed here. 
Column: 5,6, USNO B1.0 magnitudes (Monet, \et\, 2003),
Column 7: new(*) or archival $z$, Column 8: classification (see Table 1).}
\tablewidth{0pt}

\tablehead{
\colhead{ID} & \colhead{FoM} & \colhead{$\alpha$} & \colhead{$\delta$}&\colhead{R2}& \colhead{B2} & \colhead{z} & \colhead{Type}
\\
\colhead{} & \colhead{} &\multicolumn{2}{c}{(J2000 Coordinates)}
&\multicolumn{2}{c}{Magnitude}  &\colhead{}  &\colhead{} }

\startdata

\tableline
J0001+1914&0.29&00 01 08.623&+19 14 33.82&20.50&21.19&3.100&f\\ \tableline 

J0003+2129&1.69&00 03 19.348&+21 29 44.42&19.75&21.10&0.450*&f\\ \tableline

J0004+2019&1.17&00 04 35.757&+20 19 42.25&20.25&20.81&0.677*&f\\ \tableline

J0004+4615&0.28&00 04 16.128&+46 15 17.96&20.44&20.53&1.810*&f\\ \tableline

J0113+0222&3.55&01 13 43.145&+02 22 17.32&11.22&11.55&0.047&R \\ \tableline

J0122+0310&0.59&01 22 01.911&+03 10 02.43&20.07&20.91&4.000*&f \\ \tableline 

J0213+1213&0.53&02 13 05.184&+12 13 10.90&19.75&21.15&0.252:*&b \\ \tableline 


J0223+4259&0.90&02 23 11.407&+42 59 31.43&-- &--&0.021&R \\ \tableline

J0242+1742&1.64&02 42 24.268&+17 42 58.85&20.28&21.31&0.551:*&f\\ \tableline

J0246+1823&0.42&02 46 11.823&+18 23 30.08&19.5&20.94&3.590*&f \\ \tableline 

J0325+2224&1.40&03 25 36.814&+22 24 00.42&19.14&20.17&2.066*&f \\ \tableline 

J0406+0637&1.11&04 06 34.308&+06 37 14.97&19.47&19.35&0.666*&f \\ \tableline 

J0407+0742&3.01&04 07 29.087&+07 42 07.45&17.28&17.59&1.133*&b \\ \tableline 

J0409+0640&0.93&04 09 25.847&+06 40 35.09&--&--&0.805*&f\\ \tableline

J0422+1741&0.78&04 22 47.774&+17 41 15.88&19.49&19.74&0.908*&f \\ \tableline 


J0448+1127&0.54&04 48 50.413&+11 27 54.40&18.83&20.42&1.369*&f\\ \tableline

J0457+0645&0.30&04 57 07.710&+06 45 07.27&18.14&19.33&0.405&f\\ \tableline 

J0503+3403&0.59&05 03 56.786&+34 03 28.14&17.30&18.99&0.149*&R?\\ \tableline

J0505+0459&0.81&05 05 23.187&+04 59 42.73&17.58&17.24&0.954&f\\ \tableline 

J0514+5602&0.41&05 14 18.698&+56 02 11.05&--&--&2.190*&f\\ \tableline

J0533+4822&4.23&05 33 15.864&+48 22 52.82&18.25&20.34&1.160*&f\\ \tableline

J0540+2507&0.63&05 40 14.345&+25 07 55.35&--&--&0.623:*&f\\ \tableline


J0809+5218&1.24&08 09 49.189&+52 18 58.25&14.54&15.59&0.138&b \\ \tableline

J1132+0034&0.71&11 32 45.619&+00 34 27.82&17.23&17.56&--*&b\\ \tableline

J1133+0015&0.42&11 33 03.029&+00 15 48.99&18.89&18.06&1.173&f\\ \tableline 

J1133+0040&4.07&11 33 20.058&+00 40 52.84&18.88&19.43&1.633*&f\\ \tableline

J1221+4411&0.49&12 21 27.045&+44 11 29.67&17.75&18.51&1.345&f\\ \tableline 

J1224+4335&0.91&12 24 51.507&+43 35 19.28&20.22&19.71&1.872*&f\\ \tableline

J1231+0418&1.34&12 31 27.582&+04 18 01.89&17.73&18.08&1.028*&f\\ \tableline


J1321+2216&0.54&13 21 11.204&+22 16 12.10&19.37&19.47&0.943*&f\\ \tableline

J1331+1712&0.45&13 31 33.446&+17 12 50.62&18.34&18.47&--*&b\\ \tableline

J1343+2844&0.34&13 43 00.180&+28 44 07.49&16.75&17.09&0.908&f\\ \tableline 

J1419+3821&0.90&14 19 46.616&+38 21 48.49&19.25&19.33&1.832&f\\ \tableline 

J1420+3721&0.25&14 20 00.342&+37 21 34.68&18.23&18.30&0.969&f\\ \tableline 

J1421+3855&0.46&14 21 06.034&+38 55 22.83&17.39&17.84&0.490&f\\ \tableline 

J1426+3625&0.44&14 26 37.086&+36 25 09.58&20.27&20.77&1.091&f\\ \tableline 

J1603+1554&4.46&16 03 38.065&+15 54 02.38&12.15&13.97&0.109&f\\ \tableline 

J1722+6105&0.95&17 22 40.059&+61 05 59.80&19.42&19.25&2.058*&f \\ \tableline 

J1826+0149&1.39&18 26 25.066&+01 49 40.12&--&--&1.771*&f \\ \tableline 

J1826+3431&1.11&18 26 59.982&+34 31 14.10&16.20&17.66&1.814&f\\ \tableline 

J2025+3343&2.85&20 25 10.940&+33 43 00.21&--&--&0.219:*&f\\ \tableline

J2031+1219&0.26&20 31 54.999&+12 19 41.34&17.42&--&1.215&f\\ \tableline 

J2034+1154&0.61&20 34 37.110&+11 54 31.38&--&17.01&0.607*&f \\ \tableline 

J2208+6519&0.41&22 08 03.103&+65 19 38.78&--&--&1.120:*&f\\ \tableline

J2212+2355&4.11&22 12 05.970&+23 55 40.59&19.69&20.66&1.125*&f\\ \tableline 

\enddata
\end{deluxetable}


\begin{thebibliography}{}
\bibitem[Blades, \et (1980)]{bla80s}Blades, J.C., Murdoch, H.S. \& Hunstead, R.W.
1980, \mnras, 191 61.
\bibitem[Condon, \et \, (1991)]{con91}Condon, J.J., \et\, 1991, \aj, 102, 
2041
\bibitem[Fey \& Charlot (2000)]{fey00}Fey, A.L. \& Charlot, P. 2000, \apjs, 128, 17
\bibitem[Gehrels, \et (2000)]{get00}Gehrels, N., \et 2000, 
{\it Nature}, 404, 363
\bibitem[Grenier (2000)]{gre00}Grenier, I.A. 2000, \aap, 364, L93
\bibitem[Halpern \et(2001)]{hal01} Halpern, J.P., Eracleous, M, Mukherjee, R., Gotthelf, E.V. 2001, \apj, 551, 1016
\bibitem[Halpern \et (2001b)]{hal01b} Halpern, J.P.,\et  2001, \apj, 552, L125
\bibitem[Halpern \et (2002)]{hal02} Halpern, J.P., Gotthelf, E.V.,
Mirabal, N. \& Camilo, F. 2002, \apjl, 573, L41
\bibitem[Halpern, Eracleous \& Mattox (2003)]{hem03}Halpern, J.P., Eracleous, M.
\& Mattox, J.R. 2003, \aj, in press.
\bibitem[Hartman \et(1999)]{har99} Hartman, R.C., \et  1999, \apjs, 123, 79
\bibitem[Hill \et(1998)]{hil98} Hill, G.J., Nicklas, H.E., MacQueen, P.J., Tejada, C., Cobos Duenas, F.J. \& Mitsch, W. 1998 \procspie, 3355, 375
\bibitem[Jackson, \et \, (1993)]{jac93}Jackson, N., Sparks, W.B., Miley, G.K.
\& Machetto, F. 1993, \aap, 269, 128
\bibitem[Kaaret \& Cottam (1996)]{kc96}Kaaret, P. \& Cottam, J. 1996, \apjl, 462, 35 
\bibitem[Kaaret (2001)]{kar01}Kaaret, P. 2001, {\it Ast.Sp.Sci.Lib}, 267, 191 
\bibitem[Kuiper \et\, (2000)]{kui00}Kuiper, L. \et\, 2000, \aap, 359, 615.
\bibitem[Landt, \et\, (2001)]{let01}Landt, H., \et\, 2001, \mnras, 323, 757.
\bibitem[Marcha \et\, (1996)]{mar96}Marcha, M.J.M., Browne, I.W.A., 
Impey, C.D. \& Smith, P.S. 1996, \mnras, 281, 425
\bibitem[Mattox \et\,(1996)]{mat96} Mattox, J.R., \et\, 1996, \apj, 461, 396
\bibitem[Mattox \et\,(2001)]{mat01} Mattox, J.R., Hartman, R.C., Reimer, O. 2001, \apjs, 135, 155
\bibitem[Monet \et, (2003)]{mon03} Monet, D.G \et\, 2003, AJ, 125, 984
\bibitem[Mukherjee \et (2002)]{muk02}Mukherjee, R., Halpern, J., Mirabal, N. 
\& Gotthelf, E.V. 2002, \apj, 574, 693
\bibitem[Neshpor \et (1998)]{nes98}Neshpor, Y.-I., \et\, 1998, {\it AstL}, 24, 134.
\bibitem[Myers \et\,(2002)]{mye02} Myers, S.T., Jackson, N.J.,Browne, I.W.A.,Bruyn, A.G., Pearson, T.J., Readhead, A.C.S., Wilkinson, P.N., Biggs, A.D., Blandford, R.D., Fassnacht, C.D., Koopmans, L.V.E., Marlow, D.R., McKean, J.P., Norbury, M.A., Phillips, P.M., Rusin, D., Shepherd, \& M.C., Sykes, C.M. 2002, astro-ph/0211073 
\bibitem[Ramsey \et\,(1998)]{ram98} Ramsey, L. W. \et\, 1998, \procspie, 3352, 34
\bibitem[Reimer \et\,(2001)]{rei01} Reimer, O. \et\, 2001, \mnras, 324, 772
\bibitem[Roberts \et, (2002)]{rob02} Roberts, M.S.E. \et, 2002, \apjl, 577, 19
\bibitem[Roberts, Romani \& Kawai (2001)]{rrk01}Roberts, M.S.E., Romani, R.W. \&
Kawai, N. 2001, \apjs, 133, 451
\bibitem[Urry \& Padovani (1995)]{up95}Urry, C.M. \& Padovani, P. 1995, 
\pasp, 107, 803
	\bibitem[Urry (1999)]{u99}Urry, C.M., 1999, {\it Astpart. Phys.}, 11, 159.
\bibitem[Veron-Cetty \et(2001)]{ver01} Veron-Cetty, M.P., Veron, P. 2001, \aap, 374, 92
\bibitem[Wallace \et(2002)]{wal02} Wallace, P.M., Halpern, J.P., Magalhaes, A.M., Thompson, D.J. 2002, \apj, 569, 36
\bibitem[White \& Becker (1992)]{wb92}White, R.L. \& Becker, R.H. 1992, \apjs, 79, 331.
\bibitem[Yadigaroglu \& Romani (1997)]{yr97}Yadigaroglu, I.-A. \& Romani, R.W. 
1997, \apj, 476, 347.

\end{thebibliography}
\end{document}